\documentclass[12pt,onecolumn,notitlepage,nofootinbib,floatfix]{revtex4-2}
\makeatletter
\AtBeginDocument{\let\LS@rot\@undefined}
\makeatother
\usepackage{amssymb,amsmath,amsfonts} 
\usepackage{bm}
\usepackage{color} 
\usepackage{subfigure}
\usepackage{graphicx}
\usepackage[margin=1in]{geometry}
\usepackage{setspace}
\usepackage{pdfpages}
\allowdisplaybreaks

\usepackage{graphicx}
\usepackage{amsmath}
\usepackage{color}
\usepackage[normalem]{ulem}
\usepackage{multirow}
\usepackage[bookmarks=false]{hyperref} 
\hypersetup{pdfstartview=FitH,pdfhighlight=/O,colorlinks=true}

\DeclareRobustCommand\bblash{\btt{\@backslashchar}} \makeatother

\begin{document}
	
	\title{ Relativistic Virial Theorem, Limiting Compactness, and the end state of gravitational collapse }
	
	\author{Naresh Dadhich}
	\email{nkd@iucaa.in}
	\affiliation{Inter-University Centre for Astronomy \& Astrophysics,\\ Post Bag 4, Pune, 411 007, India}
	
	\affiliation{Astrophysics Research Centre, School of Mathematics, Statistics and Computer Science, University of KwaZulu-Natal, Private Bag X54001, Durban 4000, South Africa}

	
	
	\date{\today}

\begin{abstract} 

It is noteworthy that limiting compactness of a static bounded configuration is characterized by a general principle: \textit{one, by equipartition of mass between inside and outside, and the other by vanishing of energy inside.} The former implies gravitational energy being half of mass leading to limiting compactness $M/R = 4/9$ of Buchdahl star while for the latter, the two are equal giving $M/R = 1/2$ of black hole with horizon. \emph{This is the relativistic Virial theorem respectively for massive and massless particles.} It is remarkable that it prescribes that there can exist only two equilibrium states which also define limiting compactness of the object. Consequently, it leads to a profound prediction that the ultimate endproduct of gravitational collapse could only be one of the two,  Buchdahl star or black hole.


\centering
\noindent {\it  Essay written for the Gravity Research Foundation 2025 Awards for
	Essays on Gravitation}
\end{abstract} 
\maketitle

\newpage

\section{Introduction}

It is a natural question to ask, how compact can an object like a star be, does there exist a limit on its compactness? Such a question was first asked in 1959 by Hans Buchdahl, and he obtained the famous bound \cite{buch}, that goes by his name, viz., $M/R \leq 4/9$ where $M$ is the mass and $R$ the radius. He obtained this under very general conditions of pressure being isotropic, energy density being nonincreasing outward, and at the boundary, the interior metric is matched to the unique Schwarzschild exterior vacuum metric.

A Buchdahl star is defined when the bound is saturated; i.e., $M/R = 4/9$ \cite{maxforce}, and it is the limiting compactness for an object without horizon, while the ultimate limit is of a black hole, $M/R = 1/2$ with a horizon. These are the only two objects of limiting compactness.

Another important question that arises, what kind of matter configurations would such limiting objects have? If the limit is to be free of all conditionalities of matter properties like pressure degeneracy and equation of state etc, the configuration should consist of free particles interacting only through gravity, and nothing else. That is, all matter fields have already been won over by gravity leaving motion alone to fend for itself leading to the celebrated Virial equilibrium. There is, however, a critical difference from the classical Virial configuration in that it also prescribes the limiting compactness.

This is quite in contrast to the Newtonian situation where the Virial equilibrium does not define compactness. What is required is that average kinetic energy is half of average potential energy, and there is no bound on compactness $M/R$. The only bound on kinetic energy comes from the requirement that the particle should not attain the escape velocity to run off; i.e., kinetic energy should be less than potential energy which is anyway so for the Virial equilibrium. Thus, classically, Virial configurations could have arbitrary compactness. The situation is totally different for the relativistic 
case.  

It is envisioned that at infinity, the system is in an infinitely dispersed state of zero compactness having bare ADM mass $M$, and as it collapses under its own gravity, it picks up gravitational energy, $E_g(R)$, that lies in the exterior. At any given radius $R$, energy inside and outside is distributed as $E_{in}(R) = M - E_g(R)$ and $E_{out}(R) = E_g(R)$. A natural equilibrium without any conditionality could indeed be given by the equality of energy inside and outside; i.e., $M - E_g = E_g$, implying the Virial-like relation, $E_g = 1/2\; M$. Note that $E_g$ in the interior is the measure of the internal energy which in the case of free particles in motion is entirely kinetic energy. This is the relativistic Virial theorem \footnote{There have been attempts to formulate Virial theorem in GR \cite{GB94} including the one fully geometric expression \cite{DGH25}, the spirit of anchoring the whole argument here on the balance of gravitational and non-gravitational energy is rather novel leading to a neat and insightful statement. } pronouncing, kinetic energy is half of the mass of the object. Unlike the classical Virial theorem, it is remarkable that it also prescribes the limiting compactness, $M/R  = 4/9$ for massive particles with velocity, $v^2 = 8/9$, making up Buchdahl star without horizon. On the other hand, for massless particles with $v^2 = 1$, $E_{in} = M - E_g = 0$; i.e., $M = E_g$ leading to the ultimate compactness, $M/R = 1/2$, of black hole with horizon. 

We thus arrive at a very remarkable prediction that there can occur only two limiting compactness configurations, one without and the other with the horizon, and they are identified with the relativistic Virial equilibrium. In general relativity (GR), Virial equilibrium, thus, also determines the limiting compactness.

As gravitational collapse proceeds,  tidal deformations become stronger and stronger, and it is then conceivable that fluid may start breaking up into free elements turning into a Virial distribution. That is, gravitational collapse ultimately leads to the Virialization (breaking into free elements) of accreting matter. It can then only end up in one of the two available equilibria of limiting compactness, of Buchdahl star without horizon and of black hole with horizon.

This scenario leaves no room open for the occurrence of a naked singularity. We thus seem to answer in the affirmative the important question of Penrose's Weak Cosmic Censorship Conjecture \cite{Pen69}.

The key ingredient required to complete the story is the computation of gravitational energy. In GR, energy is, in general, and more so gravitational energy, is rather illdefined concept as it defies a covariant definition. This is because gravitational energy resides in space curvature that extends from the object's boundary to infinity and hence is non-localizable. There is no unique way to quantify it and therefore exist many prescriptions. The ultimate test of validity is the fact whether it yields the expected Newtonian result in the first approximation and conjures well with the overall physical perspective.

This is what we shall take up in the next section and also establish the relativistic Virial theorem. We will conclude with a discussion highlighting the overall import of the discourse.

\section{Gravitational Energy and Relativistic Virial Theorem} 

We shall here resort to the Brown-York prescription of quasi-local energy \cite{bro-yor}, which defines the energy enclosed inside radius $R$ as given by   
\begin{equation}
E(R)= \frac{1}{8\pi} \int{d^2x\sqrt{q}(k-k_0)} 
\end{equation}
where $k$ and $q$ are respectively trace of extrinsic curvature and determinant of the metric, $q_{ab}$ on $2$-surface. The reference extrinsic curvature, $k_0$ is of some reference spacetime, which for an asymptotically flat case would naturally be the Minkowski flat. This is the measure of total energy contained inside a sphere of radius $R$ around a static object.  The evaluation of the above integral for the Schwarzschild vacuum metric yields,
\begin{equation}
E(R) = R - \sqrt{R^2 - 2MR },
\end{equation}
which expands for large $R$ to give $M - (-M^2/2R) = M + M^2/2R$. It includes the desired Newtonian gravitational energy, $-M^2/2R$ in the first approximation. 

Recall that $M$ is energy at infinity and then on collapse under its own gravity, it picks up gravitational energy, $E_g(R)$ lying in the exterior, which will be given by $E(R) - M$. So we write,  
\begin{equation}
E_g(R) = E(R) - M = R - \sqrt{R^2 - 2MR } -M. 
\end{equation}

Recall that $E_{in} = M -E_g$ and $E_{out} = E_g$, the two Virial equilibrium states are then defined as follows:

\begin{equation}
E_{in} = E_{out}; \; i.e., \; E_g = 1/2\; M 
\end{equation}
for timelike particles, and 
\begin{equation}
E_{in} = 0; \; i.e.,\;  E_g = M 
\end{equation}
for null particles. \\

\emph{Relativistic Virial Theorem} then states that gravitational energy is half of the mass for the timelike equilibrium while it is equal to the mass for the null equilibrium. That is, it defines the two equilibrium states, $E_g = 1/2\; M$  and $E_g = M$ which respectively prescribe limiting compactness 
\begin{equation}
M/R = 4/9 
\end{equation}
with a timelike boundary, and 
 \begin{equation}
M/R = 1/2 
\end{equation}
with a null boundary. The former defines the limiting compactness of a Buchdahl star without a horizon, while the latter the ultimate compactness of a black hole with a horizon.

These are the only two limiting compact configurations allowed by the relativistic Virial theorem. The remarkable feature of it is that it not only determines the equilibrium states but also the limiting compactness, one without a horizon and the other with a horizon.

In the classical limit, gravitational energy, which is the measure of kinetic energy, goes as $M^2/2R$, which is half of the potential energy. What it indicates is the fact that a free particles configuration confined in a radius $R$ is always in the classical Virial equilibrium and can have arbitrary compactness.

In GR, gravity is not an external force derived from an externally given potential but instead, it is inherent in the spacetime metric. That is why kinetic and potential energy cannot be separated out. The only two parameters that are separated are non-gravitational energy, the mass,  and the gravitational energy. It is the interplay between them that determines not only the Virial equilibrium but also the limiting compactness bound. This is because gravitational energy is the function of $M$ and $R$, and then the equilibrium condition determines the compactness $M/R$.


Thus, relativistic Virial theorem is more restrictive as it allows only two equilibrium states with the prescribed limiting compactness.

Before we go any further, let us recall that there is a vast and rich literature on Buchdahl bound and its derivation in various situations which we would not take up here except to point to Ref. \cite{nd22}, and the references given in there. For the first time, the black hole horizon was defined by the equality of gravitational and non-gravitational energy in \cite{dad97}. It was then insightfully argued that at the horizon timelike particle that feels mass through gravitational pull tends to be null, that feels only gravitational energy through space curvature \cite{dad12}. There should hence be equipartition between gravitational and non-gravitational energy (mass) at the horizon.

In the context of compactness, it was first found in \cite{dad19} that when gravitational energy is half of the mass (non-gravitational energy), it defines Buchdahl bound in saturation, i.e., Buchdahl star, $M/R = 4/9$. The present elucidation is the culmination of the search for an insightful realization leading to the formulation of the relativistic Virial theorem with amazing and illuminating revelations.

\section{Discussion} 

For a fluid, limiting compactness will be indicated by the condition of incompressibility; i.e., energy density is constant. A constant density sphere is described by the unique Schwarzschild interior solution, and for that pressure at the center to remain finite requires $M/R \leq 4/9$. The bound is saturated only when pressure diverges at the center. It is then clear that no fluid distribution can attain Buchdahl star compactness, $M/R = 4/9$. This means as the configuration becomes more and more compact, the fluid state cannot be sustained due to tidal deformation and it breaks into free elements interacting only through gravity, giving rise to a Virial distribution. It is perhaps realized via Vlasov kinetic matter \cite{And-Rei06, And11}.

A considerable amount of effort has been invested in solving Einstein-Vlasov equation for building models with kinetic matter \cite{And-Rei06, And11}. In particular, configurations consisting of thin shells of particles in motion have been investigated numerically \cite{Ame-And-Log16, Ame-And-Log19}. It turns out that as the limiting compactness $M/R \to 4/9$ is approached, the shell becomes infinitely thin with $\rho, p_r, p_t$ all diverging, however,  in the limit, $p_r/\rho \to 0$ and $2p_t/\rho \to 1$. As expected, particle velocity attains the value, $v^2 = 8/9$. All these are, however, numerical results.

Buchdahl star may therefore be envisaged as made up of free particles moving with velocity, $v^2 = 8/9$, in a very thin shell at the timelike boundary. With the same force of argument, it could be said that a black hole has photons confined to a thin shell defining the null boundary of black hole -- the horizon. The latter case has not yet been possible to probe because numerical codes crash as $v^2 \to 1$.

In terms of energy balance, there are three possible limiting states: (a) $E_{out} = E_g = 0$ at infinity, it defines the limit of zero compactness of an infinitely dispersed system of bare mass $M > 0$. (b) The other limiting case is its dual,  $E_{in} = M - E_g = 0$; i.e, $E_g = M$  that defines the absolute compactness of a black hole with a horizon. Note that black hole horizon is characterized by the equality of positive energy, $M$ and gravitational energy, which is negative. (c) Lastly, $E_{in} = E_{out}$ implies $E_g = 1/2 \; M$, which defines the limiting compactness of Buchdahl star without horizon.

It is interesting to note that case (a) of zero compactness at infinity indicating zero of gravitational energy is dual to case (b) of absolute compactness when energy inside vanishes marking the equality of positive (mass) and negative (gravitational energy) energy, while case (c) lies in between the two extremes characterized by the equality of energy inside and outside. This is the relativistic Virial theorem, which not only describes equilibrium but also prescribes limiting compactness with and without horizon.

It is interesting that a Buchdahl star provides an excellent example of an object that is almost as compact as a black hole yet has timelike boundary so as to be in communication with the outside world. This was precisely what was envisaged in the proposals of the membrane paradigm \cite{mac-tho82, pri-tho86} and of the stretched horizon \cite{sus93} by invoking the existence of a fiducial timelike surface very close to the black hole horizon. There is no need for such an ad-hoc artificial construction as Buchdahl star beautifully provides all that was asked for. In that, we have a real black hole like astrophysical object without a horizon, whose equilibrium is governed by the relativistic Virial theorem. Note that, it is also as natural an endproduct of gravitational collapse as the black hole itself. It goes without saying that it is not only an excellent and natural black hole mimicker but could also perhaps be a competing candidate as an ultimate endproduct of the gravitational collapse! 

It is therefore pertinent to study some of the black hole phenomena for Buchdahl star. The most interesting atrophysical phenomenon is of energy extraction from rotating black hole via the Penrose process \cite{Pen69} and its most efficient magnetic version \cite{wag-dhu-dad85}. The magnetic Penrose process has been studied for the rotating Buchdahl star \footnote{A static object is described by the same metric whether it is a black hole or not. This is not the case for the axially symmetric Kerr solution which can only describe a rotating black hole and not any other rotating object. We employ the Kerr metric for the rotating Buchdahl star as an approximation on the ground that it is almost as compact as a black hole.} \cite{SDT24} within the limitation of employing the Kerr metric which is an approximation. Like black hole, it is also shown that non-extremal Buchdahl star cannot be extremalized \cite{SD23}, and further, it is shown to obey the weak cosmic censorship conjecture \cite{SD23b}.

It is an insightful revelation leading to the profound prediction that the relativistic Virial theorem allows only two limiting compactness equilibrium states, one of Buchdahl star without horizon and the other of black hole with horizon. All gravitationally collapsing matter has ultimately to accord to this dictum of ending at the Virial equilibrium with or without horizon.

The scenario that now emerges for generic gravitational collapse is as follows: As it proceeds and as matter fields get won over by gravity resulting in a Virial-like distribution of free elements interacting only through gravity. Then onward, the evolution is taken over by the relativistic Virial theorem allowing only two equilibrium states, one $E_g = 1/2\; M$ for massive particles resulting in the limiting compactness $M/R = 4/9$ of Buchdahl star without horizon and the other for massless particles, $E_g = M$ giving the absolute compactness $M/R = 1/2$ of black hole with horizon. The ultimate endproduct of gravitational collapse is thus either a Buchdahl star or a black hole. This is the Virial dictum that has to be obeyed.

Since Virial equilibrium defines a stable and ultimate state, the infalling matter has therefore to undergo Virialization process of breaking into free elements. This process will be driven by tidal forces resulting into production of heat flux which will flow out as electromagnetic Vaidya radiation from the boundary of the accreting zone \cite{dad-gos24}. This is a new phenomenon that collapsing object has to give out  Vaidya radiation so as to end on one of the two available equilibrium states.

It has recently been shown that an accreting black hole has first to Vaidya radiate so as the infalling matter to be in consonance with the null fluid on the horizon \cite{dad-gos24}. It also does a signal service of keeping the marginally outer trapped surface coincident with the null horizon for an accreting black hole, and thereby paving the way for Hawking radiation to propagate out. What we now have is a much more general situation where generic collapse has to get Virialized so as to posit on one of the two available equilibria states; Buchdahl star without horizon or black hole with horizon. From this follows a remarkable new prediction that gravitationally collapsing configuration as it approaches the limiting compactness, it has to Vaidya radiate. It is significant that all this is dictated by the relativistic Virial theorem.

It may be noted that the Virial equilibrium is stable by conception as it is specified by the energy balance of non-gravitational energy, mass,  and the gravitational energy. The two are organically and intimately related as the latter is caused by the former, and hence there is an in situ natural restoring mechanism for absorbing the perturbations. Thus both Buchdahl star and black hole have to be stable objects.

Now the question arises, are these two equilibria states, Buchdahl star and black hole, the natural end state of generic gravitational collapse? If yes, it seems to answer the long standing profound question of Penrose's Weak Cosmic Censorship Conjecture \cite{Pen69}. This is because as collapse proceeds it also leads to Virialization of the infalling matter and then its further evolution is entirely governed by the Virial theorem. That prescribes the unique two equilibrium states, and hence the ultimate end result has to be one of these two states. This leaves no room for the occurrence of naked singularity thereby establishing the said conjecture in the affirmative. This is by all accounts a very important result as it answers and establishes the long standing conjecture of great significance.

What has been observed \cite{bh-dis} is a compact object which could be a black hole. The question will, however, remain open until there is clear evidence of the occurrence of the horizon. It could therefore very well be a Buchdahl star which is almost as compact with $M/R = 4/9$ as against $M/R = 1/2$ of the black hole.

The important question that now arises is: \emph{Could it be the case that the ultimate endproduct of collapse is indeed a Bucchdahl star rather than a black hole?} We know that a collapsing system will first encounter the Buchdahl equilibrium and then that of the black hole. It is then clear that the second could be accessed only when the first is unstable. The key question then is the stability of the Buchdahl star, which though appears so on the general physical grounds, has, however, to be rigorously established.

If that be the case, it would be warmly welcomed as that is what was being sought for a long while, a compact object with a timelike boundary that could be in communication with the outside world as against the black hole horizon blocking out interior information entirely and absolutely. That was precisely the reason for the stretched horizon \cite{sus93} and the membrane paradigm \cite{mac-tho82, pri-tho86}
 proposals. On the other hand, very important and insightful black hole physics, astrophysics, and thermodynamics have been developed giving rise to among others, the amazing new phenomenon of Hawking radiation. Many of these may critically hinge on the existence of a null horizon.

At any rate, it opens up a new vista of interesting and exciting probe and exploration. If Buchdahl star turns out to be the ultimate end state, then the important question would be, how to carry the insightful and valuable fruits of black hole physics forward to the Buchdahl star which is almost black hole like but not quite. In this context, it may be perhaps worth remembering the dictum, in the real world, things are always "almost" and never "exact".

It is fascinating that the simple energy balance principle for inside and outside of the object leads to such profound understanding and far-reaching consequences. The principal motivation of this essay was to raise some interesting and probing questions on the important phenomenon of the ultimate end state of gravitational collapse by employing some general physical concepts and principles. To sum up let's recount that it is the relativistic 
Virial theorem that prescribes the unique two equilibrium states, one for timelike and the other for null particle distribution, which in turn require Virialization of the collapsing matter that gives out the Vaidya radiation. Since timelike equilibrium (Buchdahl star) will first occur, and the second null equilibrium (Black hole) would be accessed only when the former is won over. If the former is stable, as is expected on general physical grounds, that should be the end state of the collapse -- Buchdahl star rather than the black hole. All this has, however, to be established by the fully relativistic hydrodynamical simulations of dynamical collapse leading to actual Virialization and then ending in forming the stable Buchdahl star. 

If that turns out to be the case, it would be the most remarkable discovery answering in affirmative the Weak Cosmic Censorship Conjecture and establishing the ultimate endproduct of gravitational collapse is not black hole but Buchdahl star with a timelike boundary. With the blackhole gone, so goes out of the window the infamous information paradox. That would open up a kind of new worldview of astrophysics and gravitation. 
\section{Acknowledgements} Over a period of time, I have benefitted by interaction and discussion with several colleagues in gaining new insight and perspective. I would like to warmly thank Rituparno Goswami with whom I also had a very fruitful and insightful collaboration, Luciano Rezzolla, Haken Andreasson, Dawood Kothawala and S. Sridhar.

\end{document}